\documentclass[american,aps,pra,reprint, superscriptaddress]{revtex4-1}
\usepackage{amsmath,amscd,amsthm,amssymb}
\usepackage{mathrsfs}
\usepackage{graphicx}
\usepackage{epsf,epstopdf}
\usepackage[all]{xy}
\usepackage[unicode=true,pdfusetitle, bookmarks=true,bookmarksnumbered=false,bookmarksopen=false, breaklinks=false,pdfborder={0 0 0},backref=false,colorlinks=false] {hyperref}
\hypersetup{colorlinks=true,linkcolor=myurlcolor,citecolor=myurlcolor,urlcolor=myurlcolor}
\usepackage{graphics,epstopdf,graphicx, amsthm, amsmath, amssymb, times, braket, color, bm}
\usepackage{cleveref}
\usepackage{caption}
\usepackage{subcaption}
\usepackage{makecell}
\usepackage{appendix}
\usepackage{xcolor}
\definecolor{myurlcolor}{rgb}{0,0,0.7}

\theoremstyle{plain}

\providecommand{\theoremname}{Theorem}
\newcommand*{\myproofname}{Proof}

\makeatother
\theoremstyle{definition}

\theoremstyle{remark}

\usepackage{geometry}
\begin{document}

\title{Quantum Algorithms for Matrix Operations of Row Addition, Row Swapping, Trace Calculation and Transpose}

\author{Yu-Hang Liu}
\affiliation{School of Science, Zhejiang University of Science and Technology, Hangzhou 310023, China}

\author{Yuan-Hong Tao}
\thanks{Corresponding author: Yuanhong Tao, taoyuanhong12@126.com}
\affiliation{School of Science, Zhejiang University of Science and Technology, Hangzhou 310023, China}
\author{Jing-Run Lan}
\affiliation{School of Science, Zhejiang University of Science and Technology, Hangzhou 310023, China}
\author{Shao-Ming Fei}
\affiliation{School of Mathematical Sciences, Capital Normal University, Beijing 100048, China}

\begin{abstract}
{Quantum algorithms of matrix operations are of great significance in many fields in science and technology. In this paper, by leveraging multi-qubit Toffoli gates and basic single-qubit operations, the quantum algorithms of matrix operations of  row addition, row swapping, trace calculation and transpose are obtained. In particular, the complexities of these quantum algorithms are presented, too.}

\medskip

\end{abstract}

\maketitle

\section{Introduction}\label{Introduction}

Quantum computations offer exponential speedups over their classical counterparts by leveraging the principles of quantum mechanics. In 1980 Benioff and Manin \cite{BP} proposed the concept of quantum computing, and Feynman \cite{MYI} introduced the theoretical framework of a quantum computer in 1982. In 1985 Deutsch \cite{Deu} proposed the quantum parallel algorithm, showcasing the significant computational advantages of quantum computing over classical methods.

In 1994 Shor \cite{Sho1,Sho2} presented a quantum algorithm for prime factorization, drawing widespread attention to the advantages of quantum computing in solving certain problems. Grover \cite{Gro} followed in 1996 with a quantum search algorithm, further demonstrating quantum computing's potential in data search applications. In 2009 the HHL algorithm \cite{HHL,HHL1} was developed for solving linear systems of equations, greatly advancing the quantum algorithm researches. Several renowned quantum operations, such as quantum Fourier transform \cite{QFT1,QFT2,QFT3} and quantum phase estimation \cite{QFT3,Wang}, have significantly contributed to the field.

 Matrix operations serve as the core computational power support for quantum algorithms. The efficient quantum implementation of its basic operations (including row transformation, transposition, and trace calculation, etc.) holds significant theoretical and practical value.
In quantum chemistry simulations \cite{MYM}, the elementary row transformation of matrices can rapidly simplify the Hamiltonian matrix, reducing the time complexity of solving molecular energy spectra from the polynomial level of classical algorithms to the logarithmic level. Matrix Transpose operation enables the efficient analysis of molecular symmetry, while the calculation of the matrix trace can directly yield key physical quantities such as the number of electrons.
In quantum cryptography \cite{KSG}, elementary row transformation can optimize the encoding of quantum states to enhance security. Matrix Transpose can achieve efficient information routing in quantum communication networks, and the matrix trace can be used for the integrity detection of quantum states.
In the field of quantum machine learning \cite{PA}, elementary row transformation can accelerate the feature extraction of Quantum Support Vector Machines (QSVMs). Matrix Transpose can improve the backpropagation efficiency of quantum neural networks, and the matrix trace operation can quickly calculate kernel functions to evaluate model performance.
Given the wide-ranging influence of matrix operations in interdisciplinary fields, exploring efficient quantum computing methods to solve matrix-related algebraic problems has become a current research hotspot. 

In recent years, researchers have proposed several quantum algorithms for matrix operations. For example, the Trotter approximation formula \cite{THF} has been utilized to simulate quantum circuits, addressing complex operations such as matrix addition, multiplication and Kronecker products. Quantum algorithms for computing vector inner products \cite{ZhaoZ, SZ_2019, QZKW_arxive2022, ZQKW_arxiv2023, ZQKW_arxiv2024}  have also been developed. These algorithms leverage the superposition and entanglement of qubits and have potential applications in fields like machine learning and data processing. Based on a ``sender-receiver" quantum computing model, in 2022 Wen-Tao Qi et al. \cite{QZKW_arxive2022} proposed a class of quantum algorithms for matrix operations, including matrix-vector multiplication, matrix-matrix multiplication, matrix addition, determinant computation and matrix inversion.
In 2023 A.I. Zenchuk et al. \cite{ZQKW_arxiv2023} further developed the idea of using specific types of unitary transformations to implement linear algebra protocols according to multi-qubit Toffoli gates and the simplest single-qubit operations. Auxiliary measurements were employed to remove redundant information and extract desired results. Building on this work, in 2024 they \cite{ZQKW_arxiv2024} also introduced a method to encode the $N$ rows of a matrix into pure states of $N$ independent quantum subsystems. By selecting appropriate unitary operators, the determinant can be computed efficiently.  Furthermore, in 2025 they \cite{ZQW_arxiv2025} discussed the method of encording unnormalized matrices, and the system encoding the matrix was extended adding one more probability amplitude that satisfy the normalization condition.

However, these quantum algorithms for basic matrix operations still face challenges such as high computational complexity and low success probabilities, which limit their broader applicability.

The elementary row transformations of matrices play significant roles in solving linear systems of equations, determining matrix rank, finding invertible matrices and simplifying determinant computations, with wide-ranging value in both theoretical research and practical applications. In quantum computing, elementary row transformations can simplify matrices for efficient linear equation solving, provide theoretical support for matrix decomposition, advance large-scale matrix operations, and help to interpret the geometric meanings of linear transformations. These contributions offer new insights into the designs of quantum algorithms. Currently, there are no specialized quantum algorithms specifically designed for elementary row transformations of matrices.

Drawing on the foundational ideas from \cite{ZQKW_arxiv2023}, this paper proposes algorithms for two elementary row transformations of matrices, transpose and trace of matrices, along with their corresponding quantum circuits. These algorithms collectively involve the following key steps:

1. Construction of the initial state: matrix elements are first encoded into the probability amplitudes of quantum pure states by using tensor product operations.

2. Information separation and labeling: combining multi-qubit control operators with ancillary states, where the ancillary states are used to accurately label the useful and redundant parts of the information, achieving effective separation of the data.

3. State swaping operations: depending on the specific algorithms, C-SWAP operations are flexibly employed to precisely exchange the states of two quantum bits.

4. State superposition processing: Hadamard operators are applied to induce superposition states among quantum bits.

5. Measurement on the ancillary states: auxiliary measurement techniques are used to eliminate redundant information, ensuring that the desired results are accurately reflected in the probability amplitudes of the final state.

This paper focuses on the core algorithm layer of quantum computers and aims to design a set of quantum circuit schemes for implementing basic matrix operations. The research is carried out under the following premise: the input state, which is the quantum state encoding matrix elements, is successfully prepared. On this basis, we systematically construct a quantum circuit architecture capable of performing basic matrix operations such as row transformation transposition, and trace calculation.

This approach is of great value in the current development of quantum computing. On the one hand, quantum state preparation and quantum state measurement, as crucial aspects in quantum computing, have attracted much attention on conducting specialized studies. Transforming classical matrices into quantum pure states is a fundamental issue that needs to be addressed in the field of quantum state preparation. Moreover, converting the results of quantum state measurements into classical data interpretable by humans makes quantum readout the core part of quantum state measurement research. On the other hand, the algorithm designed in this paper fully considers the compatibility with quantum state preparation and quantum state measurement schemes, laying a solid foundation for the rapid deployment and application of quantum computers after they become practically implementable in the future.

The remainder of this paper is organized as follows: Section 2 introduces quantum algorithms for two types of elementary matrix transformations. Section 3 presents quantum algorithms for computing the trace and the transposition of a matrix. Conclusion is given in section 4.

\section{Two types of elementary row transformations}

This section presents quantum algorithms for two types of elementary row transformations of an $N\times M$ matrix $A=\{a_ {ij}\}$ ($N=2^n$, $M=2^m$), including adding one row to another and swapping two rows.

\subsection{Adding one row of a matrix to another row}

For the quantum algorithm that adds one row of a matrix to another, the quantum circuit diagram is shown in FIG.1. This algorithm requires two pure states and four auxiliary states, and is divided into seven steps.
\begin{figure}[h]
\includegraphics[width=0.5\textwidth]{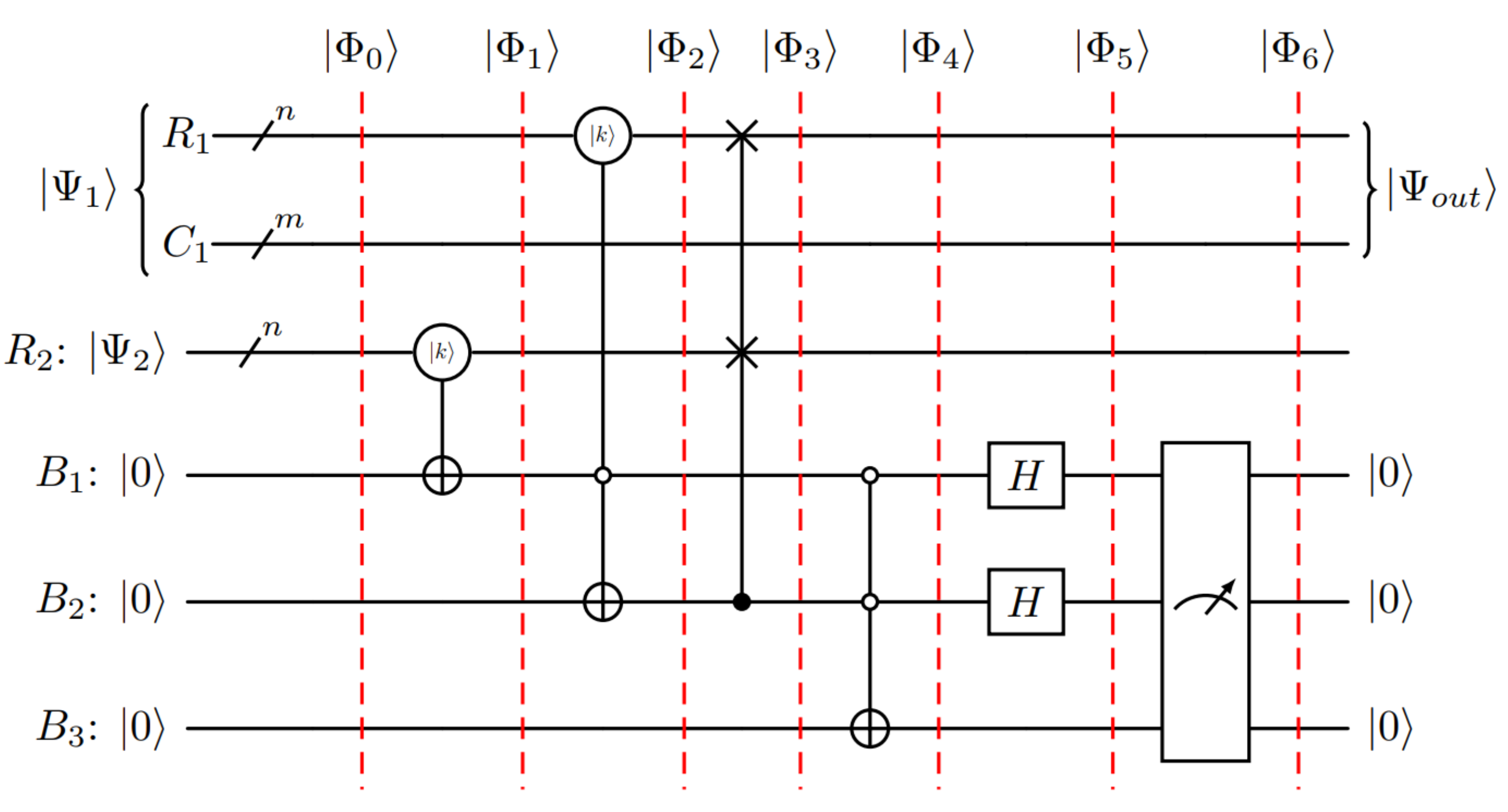}
\caption{Quantum circuit diagram for adding one row of a matrix to another row}
\end{figure}

The following steps give rise to the algorithm of adding the $(k+1)$th row of a matrix
$A$ to the $(l+1)$th row with $0\leq l,k\leq N-1$.

\textbf {Step 1: Prepare the initial state}

First, introduce an $n$-qubits register $R_1$ and an $m$-qubit register $C_1$, which are used to enumerate the rows and columns of matrix $A$, respectively. The entries $a_{ij}$ of the matrix $A$ are then encoded into a pure state as follows,
\begin{eqnarray*}\label{(1)}
|\Psi_1\rangle = \sum_{i=0}^{N-1}  \sum_{j=0}^{M-1}   a_{ij} |i\rangle_{R_1} |j\rangle_{C_1} ,\;\;~\sum_{ij}|a_{ij}|^2=1,
\end{eqnarray*}
where $|i\rangle$ is the binary representation of $i$.

Next, introduce an $N$-dimensional auxiliary column vector $Z$ (with the elements of the $k+1$ and the $l+1$ rows being $\frac{1} {\sqrt{2}}$, and all other elements being 0), along with an $n$-qubit register $R_2$. The elements of $Z$ are then encoded into pure states as follows,
\begin{eqnarray*}\label{(2)}
|\Psi_2\rangle = \frac{1}{\sqrt{2}}(|k\rangle_{R_2} +|l\rangle_{R_2}).
\end{eqnarray*}

The initial state of the entire system is then given by
\begin{eqnarray}\label{(3)}
|\Phi_0\rangle &=&  |\Psi_1\rangle \otimes |\Psi_2\rangle\\\nonumber
&=&\frac{1}{\sqrt{2}}\Big(\sum_{i=0}^{N-1} \sum_{j=0}^{M-1}  a_{ij}  |i\rangle_{R_1} |j\rangle_{C_1}  |k\rangle_{R_2}  \\\nonumber
&&
+\sum_{i=0}^{N-1} \sum_{j=0}^{M-1} a_{ij}  |i\rangle_{R_1} |j\rangle_{C_1}  |l\rangle_{R_2} \Big).
\end{eqnarray}

\textbf {Step 2: Label the two states of $R_2$.}

In order to accurately separate the useful and useless information terms in $|\Phi_0\rangle$
for subsequent calculations, we need to label the two states of $R_2$ in $|\Phi_0\rangle$ accordingly.

Introduce a 1-qubit auxiliary $B_1$ in state $|0\rangle_{B_1}$, and the following control operator,
\begin{eqnarray*}
W^{(1)}_{R_2B_1} = P_{R_2} \otimes \sigma^{(x)}_{B_1} + (I_{R_2}-P_{R_2}) \otimes I_{B_1},
\end{eqnarray*}
where $P_{R_2}=|k\rangle_{R_2} {_{R_2}}\langle k|$ is a projection operator, $\sigma^{(x)}_{B_1}$ is the standard Pauli matrix, and $I_ {B_1}$ is the identity operator acting on $B_1$. Applying the operator $W^{(1)}_{R_2B_1}$ to $|\Phi_0\rangle |0\rangle_{B_1}$, we get
\begin{eqnarray}
|\Phi_1\rangle &=&W^{(1)}_{R_2B_1} |\Phi_0\rangle \otimes|0\rangle_{B_1}\\\nonumber
&=&\frac{1}{\sqrt{2}}\Big(\sum_{i=0}^{N-1} \sum_{j=0}^{M-1} a_{ij}  |i\rangle_{R_1} |j\rangle_{C_1}  |k\rangle_{R_2} |1\rangle_{B_1} \\\nonumber
&&
+\sum_{i=0}^{N-1} \sum_{j=0}^{M-1} a_{ij}  |i\rangle_{R_1} |j\rangle_{C_1}  |l\rangle_{R_2} |0\rangle_{B_1} \Big).
\end{eqnarray}
That is, for the term where the state of $R_2$ is $|k\rangle_{R_2}$, it is labeled by
$|1\rangle_ {B_1}$, and for the term where the state of $R_2$
is $|l\rangle_{R_2}$, it is labeled by $|0\rangle_ {B_1}$.

 Kitaev A. Y. et. al  \cite{KShV} represented an important result: A control operator with $n$ control qubits can be represented by $O(n)$ Toffoli gates. Since the above control operator $W^{(1)}_{R_2B_1}$ has an $n$-qubit control register, it can be expressed by using $O(n)$ Toffoli gates. Therefore, the circuit complexity for computing $|\Phi_1\rangle$ is $O(n)=O(\log\,N)$.

\textbf {Step 3: Separate the useful information from the redundant information.}

To precisely extract the element located at the $(k+1)$th row of the second term in $|\Phi_1\rangle$, from the terms in $|\Phi_1\rangle$ labeled by $|0\rangle_ {B_1}$ we only extract the terms where $R_1$ is in the state $|k\rangle_{R_1}$.
To do so, we introduce a 1-qubit auxiliary $B_2$
in the state $|0\rangle_{B_2}$, and the following control operator,
\begin{eqnarray*}
W^{(2)}_{R_1B_1B_2} &=& P_{R_1B_1} \otimes \sigma^{(x)}_{B_2} + (I_{R_1B_1}-P_{R_1B_1}) \otimes I_{B_2},
\end{eqnarray*}
where $P_{R_1B_1}=|k\rangle_{R_1} |0\rangle_{B_1}\; {_{R_1}}\langle k| {_{B_1}}\langle 0|$ is a projection operator.

Applying the operator $W^{(2)}_{R_1B_1B_2}$ to $|\Phi_1\rangle |0\rangle_{B_2}$, we obtain
\begin{eqnarray}
|\Phi_2\rangle &=&W^{(2)}_{R_1B_1B_2} |\Phi_1\rangle \otimes |0\rangle_{B_2}\\\nonumber
&=&\frac{1}{\sqrt{2}}\Big(\sum_{i=0}^{N-1} \sum_{j=0}^{M-1}  a_{ij}  |i\rangle_{R_1} |j\rangle_{C_1}  |k\rangle_{R_2} |1\rangle_{B_1} |0\rangle_{B_2} \\\nonumber
&&+\sum_{j=0}^{M-1} a_{kj}  |k\rangle_{R_1} |j\rangle_{C_1}  |l\rangle_{R_2}  |0\rangle_{B_1} |1\rangle_{B_2} \Big)\\\nonumber
&&+|g_1\rangle_{R_1C_1R_2} |0\rangle_{B_1} |0\rangle_{B_2}.
\end{eqnarray}
That is, the terms where the state of $R_1$ and $B_1$ are $|k\rangle_{R_1}|0\rangle_ {B_1}$ are labeled by $|1\rangle_ {B_2}$, while the other terms are labeled by $|0\rangle_ {B_2}$. The states $|1\rangle_{B_1}|0\rangle_{B_2}$ and $|0\rangle_{B_1}|1\rangle_{B_2}$ in $|\Phi_2\rangle$ correspond to the useful information terms required by the algorithm. All other useless information terms are represented by $|g_1\rangle_{R_1C_1R_2}$.

The control operator $W^{(2)}_{R_1B_1B_2}$ has $n+1$ control qubits, so it can be represented in terms of $O(n)$ Toffoli gates and the circuit complexity for computing $|\Phi_2\rangle$ is $O(n)=O(\log\,N)$.

\textbf {Step 4: Use C-SWAP gate to exchange the states of $R_1$ and $R_2$ marked by $|1\rangle_ {B_2}$.}

To add the $(k+1)$th row to the $(l+1)$th row, we apply a C-SWAP gate operation to the terms in $|\Phi_2\rangle$ labeled by $|1\rangle_{B_2}$, which swaps the states of $R_1$ and $R_2$. This operation is given by the following gate,
\begin{eqnarray*}
W^{(3)}_{R_1R_2B_2}=SWAP_{R_1,R_2} \otimes |1\rangle_{B_2} \;{_{B_2}}\langle 1|
+I_{R_1R_2} \otimes |0\rangle_{B_2} \;{_{B_2}}\langle 0|.
\end{eqnarray*}
Applying $W^{(3)}_{R_1R_2B_2}$ to $|\Phi_2\rangle$ we obtain
\begin{eqnarray}
|\Phi_3\rangle &=&W^{(3)}_{R_1R_2B_2}|\Phi_2\rangle\\\nonumber
&=&\frac{1}{\sqrt{2}}\Big(\sum_{i=0}^{N-1} \sum_{j=0}^{M-1} a_{ij}  |i\rangle_{R_1} |j\rangle_{C_1}  |k\rangle_{R_2} |1\rangle_{B_1} |0\rangle_{B_2} \\\nonumber
&&
+\sum_{j=0}^{M-1} a_{kj}  |l\rangle_{R_1} |j\rangle_{C_1}  |k\rangle_{R_2}  |0\rangle_{B_1} |1\rangle_{B_2} \Big)\\\nonumber
&&
+|g_1\rangle_{R_1C_1R_2} |0\rangle_{B_1} |0\rangle_{B_2}.
\end{eqnarray}

 Note that the $SWAPs$ in the control operator $W^{(3)}_{R_1R_2B_2}$ have common single control qubit and are related to different pairs of qubits. Therefore, they can be applied simultaneously.

\textbf {Step 5: Label the useful and redundant information.}

To effectively avoid mixing useful and useless information terms in the subsequent computation process, we introduce an auxiliary qubit $B_3$ in state $|0\rangle_{B_3}$, together with the projection operator $P_ {B_1B_2}=|00\rangle_{B_1B_2} \;{_{B_1B_2}}\langle 00|$. We use $B_3$ to label the useful and useless information terms in $|\Phi_3\rangle$ accordingly by using the following control operator,
$$
W^{(4)}_{B_1B_2B_3}=P_{B_1B_2}\otimes\sigma^{(x)}_{B_3} + (I_{B_1B_2}-P_{B_1B_2}) \otimes I_{B_3}.
$$
Applying $W^{(4)}_{B_1B_2B_3}$ to $|\Phi_3\rangle|0\rangle_{B_3}$ we obtain
\begin{eqnarray}
|\Phi_4\rangle&=& W^{(4)}_{B_1B_2B_3}|\Phi_3\rangle |0\rangle_{B_3} \\\nonumber
&=&\frac{1}{\sqrt{2}}\Big(\sum_{i=0}^{N-1} \sum_{j=0}^{M-1}  a_{ij}  |i\rangle_{R_1} |j\rangle_{C_1}  |k\rangle_{R_2} |1\rangle_{B_1} |0\rangle_{B_2} \\\nonumber
  &&
+\sum_{j=0}^{M-1} a_{kj}  |l\rangle_{R_1} |j\rangle_{C_1}  |k\rangle_{R_2} |0\rangle_{B_1} |1\rangle_{B_2} \Big)|0\rangle_{B_3}\\\nonumber
&&
+|g_1\rangle_{R_1C_1R_2B_1B_2} |1\rangle_{B_3}.
\end{eqnarray}
In this way, $|0\rangle_{B_3}$ ($|1\rangle_{B_3}$) labels the useful (useless) information terms.

Since the control operator $W^{(4)}_{B_1B_2B_3}$ has two control qubit, its computational complexity is $O(1)$.

\textbf {Step 6: Use a Hadamard gate for information superposition.}

Applying the Hadamard operator $W^{(5)}_{B_1B_2}=H^{\otimes 2}$ to $B_1$ and $B_2$, we obtain
\begin{eqnarray}
|\Phi_5\rangle &=&W^{(5)}_{B_1B_2} |\Phi_4\rangle  \\\nonumber
&=&\frac{1}{(\sqrt{2})^{3}}\Big(\sum_{i=0}^{N-1} \sum_{j=0}^{M-1}  a_{ij}  |i\rangle_{R_1} |j\rangle_{C_1}
 \\\nonumber
&&+\sum_{j=0}^{M-1} a_{kj}  |l\rangle_{R_1} |j\rangle_{C_1} \Big)|k\rangle_{R_2}  |0\rangle_{B_1} |0\rangle_{B_2} |0\rangle_{B_3}\\\nonumber
&&+|g_2\rangle_{R_1C_1R_2B_1B_2B_3} .
\end{eqnarray}
Here, the terms in the state $|0\rangle_{B_1} |0\rangle_{B_2} |0\rangle_{B_3}$ are retained, which contain all the information required for further process. The remaining terms are represented as $|g_2\rangle_{R_1C_1R_2B_1B_2B_3}$.

Since only two Hadamard gates are used here, the circuit complexity is $O(1)$.

\textbf{Step 7: Measure the auxiliary state $B_1B_2B_3$ using $|000\rangle_{B_1B_2B_3}$.}

By measuring the auxiliary $B_1B_2B_3$ in the state $|000\rangle_{B_1B_2B_3}$, we get
\begin{eqnarray}
|\Phi_6\rangle &=&
|\Psi_{out}\rangle |k\rangle_{R_2},
\end{eqnarray}
where
\begin{eqnarray}
|\Psi_{out}\rangle &=&\frac{1}{G} \Big(\sum_{i\neq l}  \sum_{j=0}^{M-1}  a_{ij} |i\rangle_{R_1} |j\rangle_{C_1} \\\nonumber
&&+\sum_{j=0}^{M-1} (a_{lj}+a_{kj})  |l\rangle_{R_1} |j\rangle_{C_1}\Big)
\end{eqnarray}
with the normalization
$$
G=\big(\sum_{i\neq l} \sum_{j} |{a_{ij}}|^2+\sum_{j}|{a_{kj}+ a_{l j}}|^2\big )^{1/2}.
$$
The success probability is $\frac{G^2}{8}$. This probability clearly depends only on the matrix elements ${a_{ij}}$ and is independent of the matrix dimension.

The overall computational complexity of the algorithm is $O(n)$, mainly determined by
$W^ {(1)}_{R_2B_1}$ and $W^{(2)}_{R_1B_1B_2}$.

\subsection{Swapping two rows of a matrix}

For the quantum algorithm that swaps two rows of a matrix, the quantum circuit diagram is shown in FIG.2. The algorithm requires two pure states and four auxiliaries, and is divided into seven steps to swap the $(k+1)$th and $(l+1)$th rows of matrix $A$ ($0\leq l,k\leq N-1$).
\begin{figure}[h]
\includegraphics[width=0.5\textwidth]{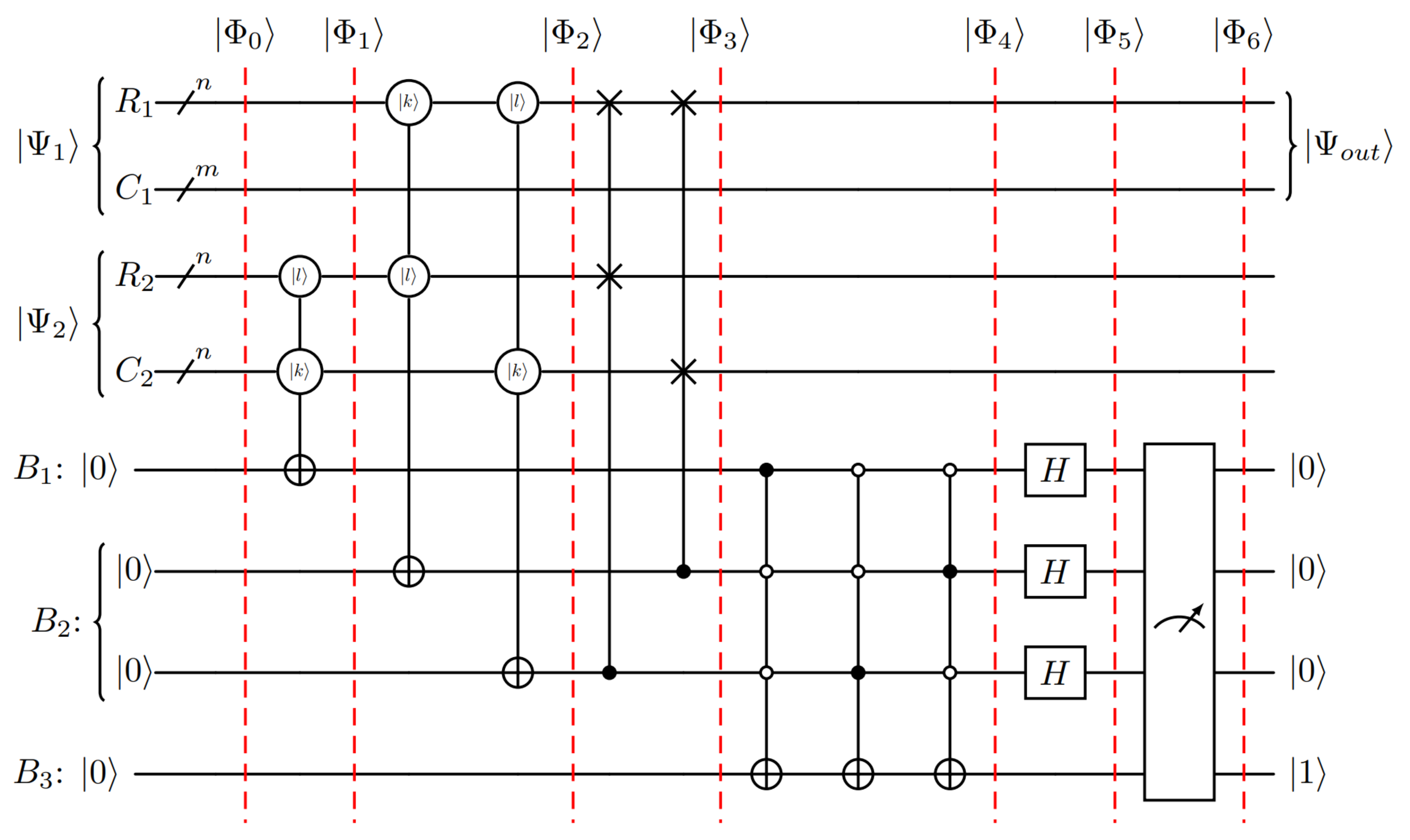}
\caption{The quantum circuit diagram for swapping two rows of a matrix.}
\end{figure}

\textbf {Step 1: Prepare the initial state.}

First, introduce an $n$-qubit register $R_1$
and an $m$-qubit register $C_1$, which enumerate the rows and columns of the matrix $A$, respectively. The elements of the matrix $A$ are then encoded into the following pure state,
\begin{eqnarray*}\label{(3)}
|\Psi_1\rangle = \sum_{i=0}^{N-1}  \sum_{j=0}^{M-1}   a_{ij} |i\rangle_{R_1} |j\rangle_{C_1} ,\;\;\sum_{ij}|a_{ij}|^2=1,
\end{eqnarray*}

Next, introduce an $N\times N$ auxiliary matrix $Z$ such that the elements in the
$(l+1)$-th row and $(k+1)$-th column, the $(k+1)$-th row and $(k+1)$-th column, and the
$(l+1)$-th row and $(l+1)$-th column are $\frac{1}{\sqrt{3}}$, the rest elements are 0; along with two $n$-qubit registers $R_2$ and $C_2$. The elements of $Z$
are then encoded into the following pure state,
\begin{eqnarray*}\label{(4)}
|\Psi_2\rangle = \frac{1}{\sqrt{3}}(|l\rangle_{R_2}|k\rangle_{C_2} +|k\rangle_{R_2}|k\rangle_{C_2}+|l\rangle_{R_2}|l\rangle_{C_2}).
\end{eqnarray*}
The initial state is of the form,
\begin{eqnarray}
|\Phi_0\rangle &=&  |\Psi_1\rangle \otimes |\Psi_2\rangle\\\nonumber
&=&\frac{1}{\sqrt{3}}\Big(\sum_{i=0}^{N-1} \sum_{j=0}^{M-1}  a_{ij}  |i\rangle_{R_1} |j\rangle_{C_1}  |l\rangle_{R_2}|k\rangle_{C_2} \\\nonumber &&+\sum_{i=0}^{N-1} \sum_{j=0}^{M-1}  a_{ij}  |i\rangle_{R_1} |j\rangle_{C_1}|k\rangle_{R_2}|k\rangle_{C_2}\\\nonumber
&&+ \sum_{i=0}^{N-1} \sum_{j=0}^{M-1}  a_{ij}  |i\rangle_{R_1} |j\rangle_{C_1}  |l\rangle_{R_2}|l\rangle_{C_2} \Big).
\end{eqnarray}

\textbf {Step 2:  Label the distinct states of $R_2$ and $C_2$.}

To accurately separate the useful from the useless information terms in $|\Phi_0\rangle$, introduce an auxiliary qubit $B_1$ in state $|0\rangle_{B_1}$, and the following control operator,
\begin{eqnarray*}
W^{(1)}_{R_2C_2B_1} = P_{R_2C_2} \otimes \sigma^{(x)}_{B_1} + (I_{R_2C_2}-P_{R_2C_2}) \otimes I_{B_1},
\end{eqnarray*}
where $P_{R_2C_2}=|l\rangle_{R_2}|k\rangle_{C_2} {_{R_2}}\langle l|{_{C_2}}\langle k|$ is the projection operator.
Applying $W^{(1)}_{R_2C_2B_1}$ to $|\Phi_0\rangle |0\rangle_{B_1}$, we get
\begin{eqnarray}
|\Phi_1\rangle &=&W^{(1)}_{R_2C_2B_1} |\Phi_0\rangle \otimes
|0\rangle_{B_1}\\\nonumber
&=&\frac{1}{\sqrt{3}}\Big(\sum_{i=0}^{N-1} \sum_{j=0}^{M-1}  a_{ij}  |i\rangle_{R_1} |j\rangle_{C_1}  |l\rangle_{R_2}|k\rangle_{C_2} |1\rangle_{B_1}\\\nonumber
&&+\sum_{i=0}^{N-1} \sum_{j=0}^{M-1}  a_{ij}  |i\rangle_{R_1} |j\rangle_{C_1}  |k\rangle_{R_2}|k\rangle_{C_2} |0\rangle_{B_1}\\\nonumber
&&+\sum_{i=0}^{N-1} \sum_{j=0}^{M-1}  a_{ij}  |i\rangle_{R_1} |j\rangle_{C_1} |l\rangle_{R_2}|l\rangle_{C_2}|0\rangle_{B_1} \Big).
\end{eqnarray}
In this way, the terms where the states of $R_2$ and $C_2$ are different are labeled by $|1\rangle_ {B_1}$, and the terms where the states of $R_2$
and $C_2$ are the same are labeled by $|0\rangle_ {B_1}$.

Since the control operator $W^{(1)}_{R_2C_2B_1}$ has $n$ control qubit, its computational complexity is $O(n)=O(\log\,N)$.

\textbf {Step 3: Separate the useful information terms from the useless information terms.}

We proceed to separate the useful information terms from the useless ones.

1) In the first sum, extract all elements except for those in the $(k+1)$th and $(l+1)$th rows.

2) In the second sum, extract all elements in the $(l+1)$th row.

3) In the third sum, extract all elements in the $(k+1)$th row.

4) For the terms in $|\Phi_1\rangle$ labeled by $|1\rangle_ {B_1}$, extract all terms except those where $R_1$ is in the states $|k\rangle_{R_1}$ and
$|l\rangle_{R_1}$.

5) For the two sums in $|\Phi_1\rangle$ labeled by $|0\rangle_ {B_1}$, extract the terms where $R_1$ is in the state $|l\rangle_ {R_1}$ and the terms where
$R_1$ is in the state $|k\rangle_{R_1}$.

We introduce an auxiliary qubit $B_2$ in state $|0\rangle_{B_2}$, and use the following control operators,
\begin{eqnarray*}
W^{(1)} &=& P_{R_1R_2} \otimes \sigma^{(x)}_{{B_2}^{1}} + (I_{R_1R_2}-P_{R_1R_2}) \otimes I_{{B_2}^{1}} ,\\\nonumber
W^{(2)} &=& P_{R_1C_2} \otimes \sigma^{(x)}_{{B_2}^{2}} + (I_{R_1C_2}-P_{R_1C_2}) \otimes I_{{B_2}^{1}} ,
\end{eqnarray*}
where $P_{R_1R_2}=|k\rangle_{R_1} |l\rangle_{R_2}\; {_{R_1}}\langle k| {_{R_2}}\langle l|$ and
$P_{R_1C_2}=|l\rangle_{R_1} |k\rangle_{C_2}\; {_{R_1}}\langle l| {_{C_2}}\langle k|$
are projection operators. $\sigma^{(x)}_{{B_2}^{i}}$ and $I_{{B_2}^{i}}$ are the operators acting on the $i$-th qubit of the auxiliary $B_2$ $(i=1,2)$.

When $W^{(2)}_{R_1R_2C_2B_2}=W^{(2)}W^{(1)}$ acts on $|\Phi_1\rangle |0\rangle_{B_2}$, we obtain
\begin{eqnarray}
|\Phi_2\rangle &=&W^{(2)}_{R_1R_2C_2B_2} |\Phi_1\rangle \otimes |0\rangle_{B_2}\\\nonumber
 &=&\frac{1}{\sqrt{3}}\Big(\sum_{i\neq l,k} \sum_{j=0}^{M-1}  a_{ij}  |i\rangle_{R_1} |j\rangle_{C_1}  |l\rangle_{R_2}|k\rangle_{C_2} |1\rangle_{B_1} |00\rangle_{B_2} \\\nonumber
 &&
+\sum_{j=0}^{M-1} a_{lj}  |l\rangle_{R_1} |j\rangle_{C_1}  |k\rangle_{R_2}|k\rangle_{C_2} |0\rangle_{B_1} |01\rangle_{B_2} \\\nonumber
&&
+\sum_{j=0}^{M-1} a_{kj}  |k\rangle_{R_1} |j\rangle_{C_1}  |l\rangle_{R_2}|l\rangle_{C_2}  |0\rangle_{B_1} |10\rangle_{B_2}\Big)\\\nonumber
&&+|g_1\rangle_{R_1C_1R_2C_2B_1B_2} .
\end{eqnarray}
That is, the terms labeled by $|1\rangle_ {B_1}|00\rangle_{B_2}$, $|0\rangle_{B_1}|01\rangle_{B_2}$
and $|0\rangle_{B_1}|10\rangle_{B_2}$ are the useful information terms required by the algorithm, while the other terms are the useless information terms denoted by $|g_1\rangle_{R_1C_1R_2C_2B_1B_2}$.

Since the control operator $W^{(2)}_{R_1R_2C_2B_2}$ has $2n$ control qubit, its computational complexity is $O(n)=O(\log\,N)$.

\textbf {Step 4: Use C-SWAP gates to swap the states of $R_1$ and $R_2$, as well as the states of $R_1$ and $C_2$.}

To swap the $(k+1)$th and $(l+1)$th rows of the matrix, we apply C-SWAP gate to the terms in $|\Phi_2\rangle$ labeled by $|10\rangle_{B_2}$ and $|01\rangle_{B_2}$, so as to swap the states of $R_1$ and $R_2$, as well as the states of $R_1$ and $C_2$. Denote
\begin{eqnarray*}
W^{(1)}&=&SWAP_{R_1,C_2} \otimes |1\rangle_{{B_2}^{1}} \;{_{{B_2}^{1}}}\langle 1|+
I_{R_1C_2} \otimes |0\rangle_{{B_2}^{1}} \;{_{{B_2}^{1}}}\langle 0|,\\\nonumber
W^{(2)}&=&SWAP_{R_1,R_2} \otimes |1\rangle_{{B_2}^{2}} \;{_{{B_2}^{2}}}\langle 1|+
I_{R_1R_2} \otimes |0\rangle_{{B_2}^{2}} \;{_{{B_2}^{2}}}\langle 0|,
\end{eqnarray*}
where $|1\rangle_{{B_2}^{1}}$ stands for that the first qubit of $B_2$ is in the state $|1\rangle$, and $|1\rangle_{{B_2}^{2}}$ for that the second qubit of $B_2$ in the state $|1\rangle$.

Applying $W^{(3)}_{R_1R_2C_2B_2}=W^{(2)}W^{(1)}$ to $|\Phi_2\rangle$, we obtain
\begin{eqnarray}
|\Phi_3\rangle &=&W^{(3)}_{R_1R_2C_2B_2}|\Phi_2\rangle\\\nonumber
&=&\frac{1}{\sqrt{3}}\Big(\sum_{i\neq l,k} \sum_{j=0}^{M-1}  a_{ij}  |i\rangle_{R_1} |j\rangle_{C_1}  |l\rangle_{R_2}|k\rangle_{C_2} |1\rangle_{B_1} |00\rangle_{B_2} \\\nonumber
&&+\sum_{j=0}^{M-1} a_{lj}  |k\rangle_{R_1} |j\rangle_{C_1}  |l\rangle_{R_2}|k\rangle_{C_2}  |0\rangle_{B_1} |01\rangle_{B_2} \\\nonumber
&&+\sum_{j=0}^{M-1} a_{kj}  |l\rangle_{R_1} |j\rangle_{C_1}  |l\rangle_{R_2}|k\rangle_{C_2} |0\rangle_{B_1} |10\rangle_{B_2}\Big)+\\\nonumber
&&+|g_2\rangle_{R_1C_1R_2C_2B_1B_2}.
\end{eqnarray}

The SWAP gates transfer the elements originally in the $(k+1)$th row to the $(l+1)$th row in the terms labeled by $|0\rangle_ {B_1}|01\rangle_{B_2}$, and the elements originally in the $(l+1)$th row to the $(k+1)$th row in the terms labeled by $|0\rangle_{B_1} |10\rangle_{B_2}$.

Note that the $SWAPs$ in the control operator $W^{(3)}_{R_1R_2B_2}$ have common single control qubit and are related to different pairs of qubits. Therefore, they can be applied simultaneously.

\textbf {Step 5: Label the useful and useless information terms.}

Introduce an auxiliary qubit $B_3$ in state $|0\rangle_{B_3}$. To use $B_3$ to label the useful and useless information terms in $|\Phi_3\rangle$, we apply the following control operator,
$$
W^{(4)}_{B_1B_2B_3}=V^{(1)}V^{(2)}V^{(3)},
$$
where $V^{(i)}=P^{(i)}_{B_1B_2}\otimes \sigma^{(x)}_{B_3} + (I_{B_1B_2}-P_{B_1B_2}) \otimes I_{B_3}$, with the projection operators
$P^{(i)}_{B_1B_2}$ $(i=1,2,3)$ given by
\begin{eqnarray*}
P^{(1)}_{B_1B_2}&=& |100\rangle_{B_1B_2} \;{_{B_1B_2}}\langle 100|,\\\nonumber
P^{(2)}_{B_1B_2}&=& |001\rangle_{B_1B_2} \;{_{B_1B_2}}\langle 001|,\\\nonumber
P^{(3)}_{B_1B_2}&=& |010\rangle_{B_1B_2} \;{_{B_1B_2}}\langle 010|.
\end{eqnarray*}

Acting $W^{(4)}_{B_1B_2B_3}$ on $|\Phi_3\rangle |0\rangle_{B_3}$, we obtain
\begin{eqnarray}
|\Phi_4\rangle&=& W^{(4)}_{B_1B_2B_3}|\Phi_3\rangle |0\rangle_{B_3} \\\nonumber
 &=&\frac{1}{\sqrt{3}}\Big(\sum_{i\neq l,k} \sum_{j=0}^{M-1}  a_{ij}  |i\rangle_{R_1} |j\rangle_{C_1}  |l\rangle_{R_2}|k\rangle_{C_2} |1\rangle_{B_1} |00\rangle_{B_2}\\\nonumber
&&
+\sum_{j=0}^{M-1} a_{lj}  |k\rangle_{R_1} |j\rangle_{C_1}  |l\rangle_{R_2}|k\rangle_{C_2} |0\rangle_{B_1} |01\rangle_{B_2} \\\nonumber
&&
+\sum_{j=0}^{M-1} a_{kj}  |l\rangle_{R_1} |j\rangle_{C_1}  |l\rangle_{R_2}|k\rangle_{C_2}  |0\rangle_{B_1} |10\rangle_{B_2}\Big) |1\rangle_{B_3} \\\nonumber
&&+
|g_2\rangle_{R_1C_1R_2C_2B_1B_2} |0\rangle_{B_3}.
\end{eqnarray}
Here, $|1\rangle_{B_3}$ labels the useful information term, while $|0\rangle_{B_3}$ labels the useless information term.

Since the control operator $W^{(4)}_{B_1B_2B_3}$ has one control qubit, its computational complexity is $O(1)$.

\textbf {Step 6: Use Hadamard gates for information superposition.}

Applying the Hadamard operator $W^{(5)}_{B_1B_2}=H^{\otimes 3}$ to $B_1$ and $B_2$, we have
\begin{eqnarray}
|\Phi_5\rangle &=&W^{(5)}_{B_1B_2} |\Phi_4\rangle  \\\nonumber
&=&\frac{1}{(\sqrt{2})^{3}\cdot\sqrt{3}}\Big(\sum_{i\neq l,k} \sum_{j=0}^{M-1}  a_{ij}  |i\rangle_{R_1} |j\rangle_{C_1}\\\nonumber
&& +\sum_{j=0}^{M-1} a_{lj}  |k\rangle_{R_1} |j\rangle_{C_1}
+\sum_{j=0}^{M-1} a_{kj}  |l\rangle_{R_1} |j\rangle_{C_1}  \Big)  \\\nonumber
&&|l\rangle_{R_2}|k\rangle_{C_2}|0\rangle_{B_1} |00\rangle_{B_2} |1\rangle_{B_3}+|g_3\rangle_{R_1C_1R_2C_2B_1B_2B_3},
\end{eqnarray}
where the terms in the state $|0\rangle_{B_1} |00\rangle_{B_2} |1\rangle_{B_3}$ are retained, the other terms are denoted by $|g_3\rangle_{R_1C_1R_2B_1B_2B_3}$.

At this point, we have successfully swaped the elements of the $(k+1)$-th row of matrix $A$ with the corresponding elements of the $(l+1)$-th row.

Since only three Hadamard gates are used here, the circuit complexity is $O (1)$.

\textbf {Step 7: Measure the auxiliary $B_1B_2B_3$ in the state $|0001\rangle_{B_1B_2B_3}$}

We measure the auxiliary $B_1B_2B_3$ in the state $|0001\rangle_{B_1B_2B_3}$ to remove the useless information terms, and obtain
\begin{eqnarray}
&&
|\Phi_6\rangle=
|\Psi_{out}\rangle |l\rangle_{R_2}|k\rangle_{C_2},
\end{eqnarray}
where
\begin{eqnarray}
 |\Psi_{out}\rangle &=& \Big(\sum_{i\neq l,k} \sum_{j=0}^{M-1}  a_{ij}  |i\rangle_{R_1} |j\rangle_{C_1}    \\\nonumber
&&
+\sum_{j=0}^{M-1} a_{lj}  |k\rangle_{R_1} |j\rangle_{C_1}+\sum_{j=0}^{M-1} a_{kj}  |l\rangle_{R_1} |j\rangle_{C_1}  \Big).
\end{eqnarray}

The success probability is {$\frac{1}{24}$}. This probability is independent of the matrix dimension.

The overall computational complexity of the algorithm is $O(n)$, mainly determined by
$W^{(1)}_{R_2C_2B_1}$ and $W^{(2)}_{R_1R_2C_2B_2}$.

For specific examples of this algorithm, please refer to Appendix 1.

\section{Matrix Trace Calculation and Transpose}

In this section we present quantum algorithms for trace calculation of an $N\times N$ matrix and transposition of an $N\times M$ matrix.

\subsection{Matrix Trace Calculation}

The quantum circuit for calculating the trace of the matrix $S$ is shown in FIG.3. Assume that a matrix $S$=$\{a_{ik}\}$ is of size $N\times N$ with $N = 2^n$. This algorithm requires one pure state and three auxiliary qubits, and is divided into the following six steps.
\begin{figure}[h]
\includegraphics[width=0.5\textwidth]{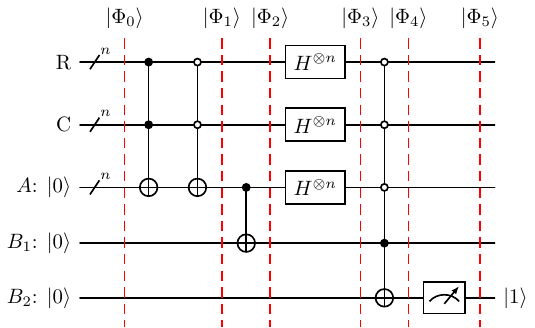}
\caption{Quantum circuit for calculating matrix trace.}
\end{figure}

\textbf {Step 1: Prepare the initial state.}

First, introduce two $n$-qubit registers $R$ and $C$, which are used to enumerate the rows and columns of the matrix $S$, and encode the elements of the matrix $S$ to obtain the following pure initial state,
\begin{eqnarray}
|\Phi_0\rangle =\sum_{i=0}^{N-1}\sum_{k=0}^{N-1} a_{ik} |i\rangle_{R}|k\rangle_{C} ,\;\; {\sum_{ik}|a_{ik}|^2=1}.
\end{eqnarray}

\textbf {Step 2: Mark the terms where the states of $R$ and $C$ are the same.}

To calculate the sum of the diagonal elements of matrix $S$, we need to mark the diagonal elements of $S$, i.e., the terms where the states of registers $R$
and $C$ are the same. To do this, introduce an $n$-qubit auxiliary $A$
in the state $|0\rangle_{A}$, and an operator
\begin{eqnarray*}
W_{j}^{(m)} &=& P^{(m)}_{j} \otimes \sigma^{(x)}_{A^{(j)}} + (I_{j}-P^{(m)}_{j}) \otimes I_{A^{(j)}},
\end{eqnarray*}
where $P^{(m)}_j=|m_{j}\rangle_{R}|m_{j}\rangle_{C}  \;{ _{R}}\langle m_{j}|  {_{C}}\langle m_{j}|$, $m=0,1$, $j=1,2,...,n$, is the projection operator acting on the $j$-th qubit of registers $R$ and $C$, $\sigma^{(x)}_{A^{(j)}}$ is the Pauli matrix acting on the
$j$-th qubit of auxiliary $A$, $I_{A^{(j)}}$ is the identity operator acting on the
$j$-th qubit of auxiliary $A$.

Applying the operator $W_{RCA}^{(1)}=\prod_{j=1}^{n}W_{j}^{0}W_{j}^{1}$ to $|\Phi_0\rangle |0\rangle_{A}$, we obtain
\begin{eqnarray}
	|\Phi_1\rangle &=& W^{(1)}_{RCA}  |\Phi_0\rangle|0\rangle_{A}\\\nonumber
	&=&\left(\sum_{i=0}^{N-1} a_{ii} |i\rangle_{R}  |i\rangle_{C}\right) |N-1\rangle_{A} +
	|g_1\rangle_{RCA}.
\end{eqnarray}

In the process of realizing the quantum state $|\Phi_1\rangle$, we check whether the
$j$-th qubits of registers $R$ and $C$ are the same. If their $j$-th qubits are the same, the
$j$-th qubit of $A$ is flipped from $|0\rangle$ to $|1\rangle$. If the states of all $n$ qubits of $R$ and $C$ are the same, the states of all $n$ qubits of $A$ are flipped from $|0\rangle_{A}$ to $|N-1\rangle_{A}$.

The control operator $W^{(1)}_{RCA}$ has $2n$ control qubits, so the circuit complexity for computing $|\Phi_1\rangle$ is $O(n)$.

\textbf {Step 3: Mark the useful and useless information terms.}

Currently, all useful information terms are stored in the first sum of $|\Phi_1 \rangle$, while the second sum $|g_1\rangle_{{RCA}}$ represents useless information terms that are not needed by the algorithm.

To separately mark the useful and useless information terms in the state $|\Phi_1\rangle$, we introduce the projection operator
$
P_A=|N-1\rangle_A\; {_A}\langle N-1|,
$
and a 1-qubit auxiliary $B_1$
in the state $|0\rangle_{B_1}$.
The control operator
$
W^{(2)}_{AB_1}=P_{A}\otimes \sigma^{(x)}_{B_1} +(I_{A}- P_{A})\otimes  I_{B_1}
$
is then applied to $A$ and $B_1$, giving rise to
\begin{eqnarray}
|\Phi_2\rangle &=& W^{(2)}_{AB_1}|\Phi_1\rangle |0\rangle_{B_1} \\\nonumber
&=&\left(\sum_{i=0}^{N-1} a_{ii} |i\rangle_{R}  |i\rangle_{C}\right) \otimes
|N-1\rangle_{A}\otimes  |1\rangle_{B_1} \\\nonumber
&&+|g_1\rangle_{RCA}\otimes  |0\rangle_{B_1} .
\end{eqnarray}
Here, the useful information terms are marked by $|1\rangle_{B_1}$, and the useless information terms are marked by $|0\rangle_{B_1}$.

 Since the control operator $W^{(2)}_{AB_1}$ has $n$ control qubits, its computational complexity is $O(n)=O(\log\,N)$.

\textbf {Step 4: Use the Hadamard gate for information superposition.}

Up to this point, we have obtained the diagonal elements of
$S$. To compute the sum we apply the Hadamard operator $W^{(3)}_{RCA}=H^{\otimes 3n}$ to all qubits in $|\Phi_2\rangle $ except for the auxiliary $B_1$. We get
\begin{eqnarray}
|\Phi_3\rangle &=&W^{(3)}_{RCA}|\Phi_2\rangle\\\nonumber
& =&\frac{1}{2^{3 n/2}}
\left(\sum_{i=0}^{N-1} a_{ii} \right)|0\rangle_{R} |0\rangle_{C} |0\rangle_{A}|1\rangle_{B_1} +
|g_2\rangle_{RCAB_1}.
\end{eqnarray}
In the state $|\Phi_3\rangle$, we select the terms in the state $|0\rangle_{R} |0\rangle_{C} |0\rangle_{A}|1\rangle_{B_1}$, as these terms contain all the useful information required for the trace calculation. The other terms are represented by $|g_2\rangle_{RCAB_1}$.

Since  $3n$ Hadamard gates are used here, the circuit complexity is $O (n)$.

\textbf {Step 5: Re-mark the useful and useless information terms.}

Since new useless information terms were generated in the previous step and the useful information is stored in the first term of $|\Phi_3\rangle$, we need to re-mark the useful and useless information terms. To do this, we introduce an auxiliary qubit $B_2$ in state $|0\rangle_{B_2}$, as well as the projection operator
$
P_{RCAB_1}= |0\rangle_{R}|0\rangle_{C}|0\rangle_{A}|1\rangle_{B_1}\;
{_{R}}\langle 0| {_{C}}\langle 0| _{A}\langle 0|  _{B_1}\langle 1|.
$
We apply the following control operator
$$
W^{(4)}_{RCAB_1B_2}=P_{RCAB_1} \otimes \sigma^{(x)}_{B_2}+
(I_{RCAB_1} -P_{RCAB_1} )\otimes I_{B_2}\,
$$
to $|\Phi_3\rangle \otimes |0\rangle_{B_2}$. We obtain
\begin{eqnarray}
	|\Phi_4\rangle &=&W^{(4)}_{RCAB_1B_2}|\Phi_3\rangle |0\rangle_{B_2}
\\\nonumber
&=&\frac{1}{2^{3 n/2}}
\left(\sum_{i=0}^{N-1} a_{ii} \right)|0\rangle_{R} |0\rangle_{C} |0\rangle_{A}|1\rangle_{B_1} |1\rangle_{B_2}\\\nonumber
&&+|g_2\rangle_{RCAB_1} |0\rangle_{B_2}.
\end{eqnarray}
Here, the useful information terms are marked by $|1\rangle_{B_2}$, and the useless information terms are marked by $ |0\rangle_{B_2} $.

Since the control operator $W^{(4)}_{RCAB_1B_2}$ has $3n+1$ control qubit, its computational complexity is $O(n)$.

\textbf {Step 6: Measure the auxiliary qubit $B_2$ in the state $|1\rangle_{B_2}$.}

The trace of the matrix $S$
is stored in the quantum state $|\Phi_4\rangle$
 in the form of probability amplitudes, with the useful information marked by $|1\rangle_{B_2}$. To extract the useful information, we measure the auxiliary qubit $B_2$ in the state $|1\rangle_{B_2}$, and obtain
\begin{eqnarray}
|\Phi_5\rangle &=&\frac{\sum_{i=0}^{N-1} a_{ii}}
{| \sum_{i=0}^{N-1} a_{ii}| }
|0\rangle_{R} |0\rangle_{C} |0\rangle_{A}|1\rangle_{B_1}.
\end{eqnarray}
The trace is obtained based on the measurement probability $\frac{| \sum_{i=0}^{N-1} a_{ii}|^2} {2^{3n}}$.
This probability is related of the matrix dimension.

The overall computational complexity of the algorithm is $O(n)$, mainly determined by
$W^{(1)}_{RCA}$, $W^{(2)}_{AB_1}$, $W^{(3)}_{RCA}$ and $W^{(4)}_{RCAB_1B_2}$.

\subsection{Matrix Transpose}

The quantum circuit for matrix transpose is shown in FIG.4. Assume that a matrix $S$=$\{a_{ik}\}$ is of size $N\times M$ with $N = 2^n$ and $M=2^m$. The algorithm requires two pure states and is divided into two steps.
\begin{figure}[h]
\includegraphics[width=0.3\textwidth]{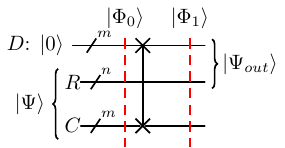}
\caption{Quantum Circuit for Matrix Transpose.
}
\end{figure}

\textbf {Step 1: Prepare the initial state.}

First, introduce an $n$-qubit register $R$ and an $m$-qubit register $C$, where $R$ and $C$ enumerate the rows and columns of the matrix $S$, respectively. The elements of matrix $S$
are encoded into the following pure state,
\begin{eqnarray*}
|\Psi\rangle =\sum_{i=0}^{N-1}\sum_{k=0}^{M-1}a_{ik}|i\rangle_{R}|k\rangle_{C},\;\;\sum_{ik}|a_{ik}|^2=1,
\end{eqnarray*}

Then, introduce a pure state $|0\rangle_{D}$ for an
$m$-qubit register $D$.
The initial state is give by
\begin{eqnarray}
|\Phi_0\rangle &=&  |0\rangle_{D} \otimes|\Psi\rangle  \\\nonumber
&=&\sum_{i=0}^{N-1}\sum_{k=0}^{M-1}a_{ik}|0\rangle_{D}|i\rangle_{R}|k\rangle_{C}.
\end{eqnarray}

\textbf {Step 2: Use a SWAP gate to exchange the states of $D$ and $C$.}

 We use a SWAP gate to exchange the states of the $D$
and $C$ registers. The operator $SWAP_{DC}$ is applied to $|\Phi_0\rangle$, yielding
\begin{eqnarray}
|\Phi_1\rangle &=&  SWAP_{DC}|\Phi_0\rangle  \\\nonumber
&=& |\Psi_{out}\rangle|0\rangle_{C},
\end{eqnarray}
where
\begin{eqnarray}
|\Psi_{out}\rangle &=& \Big(\sum_{i=0}^{N-1}\sum_{k=0}^{M-1}a_{ik}|k\rangle_{D}|i\rangle_{R}\Big).
\end{eqnarray}
Now, the new rows and new columns of the current target matrix are enumerated by the states $D$ and $R$ respectively.
Clearly, the success probability of this algorithm is 1.

Since the exchange operator $SWAP_{DC}$ only involves exchange operations when acting on different qubit pairs, its computational complexity is $O(m)$.

{\bf Remark.} Formally the non-square matrix can be made a
square matrix by including the proper number of zero rows or columns. In
this case transposition of non-square matrix can be done as a transposition
of a square matrix, i.e. by the swap operator $SWAP_{RC}$. The complexity of this operation is $O(n)$.

\section{Conclusion}

We have developed quantum algorithms for two types of elementary row transformations of matrices, the trace and transpose of matrices. The operational mechanism of all these algorithms is as follows: First, construct the initial state of the entire system. Second, utilize unitary operations and ancillary qubits to mark the information items required by the algorithm. Third, apply the Hadamard transformation to superpose these information items. Finally, measure the ancillary qubits to eliminate the useless information items and retain only the useful ones.

In all these algorithms, the final result is stored in the probability amplitude of certain quantum states. Our success probability for adding one row of a matrix to another is $\frac{G^2}{8}$. The success probability for swapping two rows of a matrix is $\frac{1}{24}$, while the success probability of matrix transpose is 1.
The success probability for the trace calculation is $\frac{| \sum_{i=0}^{N-1} a_{ii}|^2} {2^{3n}}$. The success probabilities either depend solely on the elements of the matrix itself, or are fixed or related to the dimensions of the matrix. The larger the dimension, the smaller the success probability. To address this challenge, amplitude amplification techniques may be employed. It is worth noting that the computational complexity of these algorithms is related to the dimension of the matrices. It increases logarithmically as the matrix dimension grows. Specifically, the computational complexity is $O(n)$ for elementary row operations and trace calculation, and $O(m)$ for matrix transpose.

It would be desired to develop new methods to improve the success probability given by the final measurement in the algorithm, or explore the designs of other quantum algorithms for matrix operations based on multi-qubit Toffoli type gates and the simplest single qubit operations. It would be also interesting to conduct further researches on quantum algorithms for other matrix operations.

\textbf{Acknowledgments}

This work is being supported by the following: the National Natural Science Foundation of China (NSFC) under Grants 11761073, 12075159, and 12171044; the Academician Innovation Platform of Hainan Province. We are deeply grateful to Prof. Junde Wu and Alexander I. Zenchuk for their active participation and invaluable contributions to this research.

\appendices 
\section*{Appendix 1. Swapping two rows of a matrix}
For the quantum algorithm that swaps two rows of a matrix, we will provide a detailed explanation of the quantum algorithm that swaps the 4th and 2th rows of the following $4 \times 4$  matrix $A$,
$$
 A=
\begin{pmatrix}
 \frac{1}{4} && \frac{1}{16} && \frac{3}{16} && \frac{3}{16} \\
   &&  && && \\
  0 && \frac{1}{2} && \frac{1}{8} && \frac{1}{8} \\
  &&  && && \\
  \frac{7}{16} && 0 && \frac{1}{4} && 0 \\
  &&  && && \\
  \frac{3}{16} && \frac{3}{16} && \frac{1}{8} && \frac{1}{2}
\end{pmatrix}
$$
\textbf {Step 1: Construct the initial state of the entire system.}

First, introduce two $2$-qubit register $R_1$ and $C_1$, which will enumerate the rows and columns of the matrix $A$, respectively. The elements of the matrix $A$ are then encoded into the following pure state:
\begin{eqnarray*}
|\Psi_1\rangle &=&
\frac{1}{4} |00\rangle_{R_1}|00\rangle_{C_1}+\frac{1}{16} |00\rangle_{R_1}|01\rangle_{C_1} +\frac{3}{16} |00\rangle_{R_1}|10\rangle_{C_1}\\\nonumber
&&+\frac{3}{16} |00\rangle_{R_1}|11\rangle_{C_1} +0 |01\rangle_{R_1}|00\rangle_{C_1}+\frac{1}{2} |01\rangle_{R_1}|01\rangle_{C_1}\\\nonumber
&&+\frac{1}{8} |01\rangle_{R_1}|10\rangle_{C_1}+\frac{1}{8} |01\rangle_{R_1}|11\rangle_{C_1} +\frac{7}{16} |10\rangle_{R_1}|00\rangle_{C_1}\\\nonumber
&&+0 |10\rangle_{R_1}|01\rangle_{C_1}+\frac{1}{4} |10\rangle_{R_1}|10\rangle_{C_1}+0 |10\rangle_{R_1}|11\rangle_{C_1} \\\nonumber
&&+\frac{3}{16} |11\rangle_{R_1}|00\rangle_{C_1}+\frac{3}{16} |11\rangle_{R_1}|01\rangle_{C_1}+\frac{1}{8} |11\rangle_{R_1}|10\rangle_{C_1}\\\nonumber
&&+\frac{1}{2} |11\rangle_{R_1}|11\rangle_{C_1}.
\end{eqnarray*}
Next, introduce an $2 \times 2$ auxiliary matrix $Z$,
$$
Z=
\begin{pmatrix}
 0 && 0 && 0 && 0 \\
 &&  && && \\
  0 && \frac{1}{\sqrt{3}} && 0 && \frac{1}{\sqrt{3}} \\
  &&  && && \\
  0 && 0 && 0 && 0 \\
  &&  && && \\
  0 && 0 && 0 && \frac{1}{\sqrt{3}}
\end{pmatrix},
$$
along with two
$2$-qubit registers $R_2$ and $C_2$. The elements of $Z$
are then encoded into the following pure state:
$$|\Psi_2\rangle = \frac{1}{\sqrt{3}}(|01\rangle_{R_2}|11\rangle_{C_2} +|11\rangle_{R_2}|11\rangle_{C_2} +|01\rangle_{R_2}|01\rangle_{C_2} ).$$

The two pure states $|\Psi_1\rangle$ and $|\Psi_2\rangle$
can be tensor-producted to construct the initial state of the entire system
\begin{eqnarray}
|\Phi_0\rangle &=&  |\Psi_1\rangle \otimes |\Psi_2\rangle\\\nonumber
&=&\frac{1}{\sqrt{3}}(|\Psi_1\rangle|01\rangle_{R_2}|11\rangle_{C_2}
+|\Psi_1\rangle|11\rangle_{R_2}|11\rangle_{C_2}\\\nonumber
&&+|\Psi_1\rangle|01\rangle_{R_2}|01\rangle_{C_2}
)
\end{eqnarray}

\textbf {Step 2: Label the distinct states of $R_2$ and $C_2$.}

Introduce a 1-qubit auxiliary $B_1$with the state$|0\rangle_{B_1}$, and construct the following control operator:
\begin{eqnarray*}
W^{(1)}_{R_2C_2B_1} = P_{R_2C_2} \otimes \sigma^{(x)}_{B_1} + (I_{R_2C_2}-P_{R_2C_2}) \otimes I_{B_1},
\end{eqnarray*}
where $P_{R_2C_2}=|01\rangle_{R_2}|11\rangle_{C_2} {_{R_2}}\langle 01|{_{C_2}}\langle 11|$ is the projection operator.
When the above operator acts on $|\Phi_0\rangle |0\rangle_{B_1}$, we get:
\begin{eqnarray}
|\Phi_1\rangle &=&W^{(1)}_{R_2C_2B_1} |\Phi_0\rangle \otimes
|0\rangle_{B_1}\\\nonumber
&=&\frac{1}{\sqrt{3}}(|\Psi_1\rangle|01\rangle_{R_2}|11\rangle_{C_2}|1\rangle_{B_1}\\\nonumber
&&+|\Psi_1\rangle|01\rangle_{R_2}|01\rangle_{C_2}|0\rangle_{B_1}+|\Psi_1\rangle|11\rangle_{R_2}|11\rangle_{C_2}|0\rangle_{B_1}) .
\end{eqnarray}
That is, label the terms where the states of $R_2$
and $C_2$ are different with $|1\rangle_ {B_1}$, and label the terms where the states of $R_2$
and $C_2$ are the same with $|0\rangle_ {B_1}$.

\textbf {Step 3: Separate the useful information terms from the useless information terms.}

Introducing a $1$-qubit ancilla $B_2$ with a state of $|0\rangle_{B_2}$, construct the following control operator

\begin{eqnarray*}
W^{(1)} &=& P_{R_1R_2} \otimes \sigma^{(x)}_{{B_2}^{1}} + (I_{R_1R_2}-P_{R_1R_2}) \otimes I_{{B_2}^{1}} ,\\\nonumber
W^{(2)} &=& P_{R_1C_2} \otimes \sigma^{(x)}_{{B_2}^{2}} + (I_{R_1C_2}-P_{R_1C_2}) \otimes I_{{B_2}^{1}} ,
\end{eqnarray*}
where $P_{R_1R_2}=|11\rangle_{R_1} |01\rangle_{R_2}\; {_{R_1}}\langle 11| {_{R_2}}\langle 01|$,
$P_{R_1C_2}=|01\rangle_{R_1} |11\rangle_{C_2}\; {_{R_1}}\langle 01| {_{C_2}}\langle 11|$
is the projection operator.
Apply $W^{(2)}_{R_1R_2C_2B_2}=W^{(2)}W^{(1)}$ to $|\Phi_1\rangle |0\rangle_{B_2}$, obtained:
\begin{eqnarray}
&&
|\Phi_2\rangle= W^{(2)}_{R_1R_2C_2B_2} |\Phi_1\rangle |0\rangle_{B_2} \\\nonumber
&&=\frac{1}{\sqrt{3}}\Big[\Big(\frac{1}{4}|00\rangle_{R_1}|00\rangle_{C_1}|01\rangle_{R_2}|11\rangle_{C_2}\\\nonumber
&&
+\frac{1}{16}|00\rangle_{R_1}|01\rangle_{C_1}|01\rangle_{R_2}|11\rangle_{C_2}\\\nonumber
&&+\frac{3}{16}|00\rangle_{R_1}|10\rangle_{C_1}|01\rangle_{R_2}|11\rangle_{C_2}\\\nonumber
&&
+\frac{3}{16}|00\rangle_{R_1}|11\rangle_{C_1}|01\rangle_{R_2}|11\rangle_{C_2}\\\nonumber
&&+\frac{7}{16}|10\rangle_{R_1}|00\rangle_{C_1}|01\rangle_{R_2}|11\rangle_{C_2}\\\nonumber
&&
+0 |10\rangle_{R_1}|01\rangle_{C_1}|01\rangle_{R_2}|11\rangle_{C_2}\\\nonumber
&&+\frac{1}{4}|10\rangle_{R_1}|10\rangle_{C_1}|01\rangle_{R_2}|11\rangle_{C_2}\\\nonumber
&&
+0 |10\rangle_{R_1}|11\rangle_{C_1}|01\rangle_{R_2}|11\rangle_{C_2}
\Big)|1\rangle_{B_1}|00\rangle_{B_2}\\\nonumber
&&+\Big(\frac{3}{16}|11\rangle_{R_1}|00\rangle_{C_1}|01\rangle_{R_2}|01\rangle_{C_2}\\\nonumber
&&
+\frac{3}{16}|11\rangle_{R_1}|01\rangle_{C_1}|01\rangle_{R_2}|01\rangle_{C_2}\\\nonumber
&&+\frac{2}{16}|11\rangle_{R_1}|10\rangle_{C_1}|01\rangle_{R_2}|01\rangle_{C_2}\\\nonumber
&&
+\frac{2}{4}|11\rangle_{R_1}|11\rangle_{C_1}|01\rangle_{R_2}|01\rangle_{C_2}\Big)|0\rangle_{B_1}|10\rangle_{B_2}\\\nonumber
&&+\Big(0|01\rangle_{R_1}|00\rangle_{C_1}|11\rangle_{R_2}|11\rangle_{C_2}\\\nonumber
&&
+\frac{2}{4} |01\rangle_{R_1}|01\rangle_{C_1}|11\rangle_{R_2}|11\rangle_{C_2} \\\nonumber
&&+\frac{2}{16}|01\rangle_{R_1}|10\rangle_{C_1}|11\rangle_{R_2}|11\rangle_{C_2}\\\nonumber
&&
+\frac{2}{16}|01\rangle_{R_1}|11\rangle_{C_1}|11\rangle_{R_2}|11\rangle_{C_2}\Big)|0\rangle_{B_1}|01\rangle_{B_2}\Big]\\\nonumber
&&+|g_1\rangle_{R_1C_1R_2C_2B_1B_2}
\end{eqnarray}

\textbf {Step 4: Use C-SWAP gates to swap the states of $R_1$ and $R_2$, as well as the states of $R_1$ and $C_2$.}

To achieve the goal of swapping the $4$ th and $2$ th rows of the matrix, apply C-SWAP gate operations to the terms in $|\Phi_2\rangle$
labeled by $|10\rangle_{B_2}$ and $|01\rangle_{B_2}$, swapping the states of $R_1$ and $R_2$, as well as the states of $R_1$ and $C_2$. For this, construct the following control operators:
\begin{eqnarray*}
W^{(1)}&=&SWAP_{R_1,C_2} \otimes |1\rangle_{{B_2}^{1}} \;{_{{B_2}^{1}}}\langle 1|\\\nonumber
&&
+I_{R_1C_2} \otimes |0\rangle_{{B_2}^{1}} \;{_{{B_2}^{1}}}\langle 0|\\\nonumber
W^{(2)}&=&SWAP_{R_1,R_2} \otimes |1\rangle_{{B_2}^{2}} \;{_{{B_2}^{2}}}\langle 1|\\\nonumber
&&
+I_{R_1R_2} \otimes |0\rangle_{{B_2}^{2}} \;{_{{B_2}^{2}}}\langle 0|
\end{eqnarray*}
Apply $W^{(3)}_{R_1R_2C_2B_2}=W^{(2)}W^{(1)}$ to $|\Phi_2\rangle$

\begin{eqnarray}
|\Phi_3\rangle &=&W^{(3)}_{R_1R_2C_2B_2}|\Phi_2\rangle\\\nonumber
&=&\frac{1}{\sqrt{3}}\Big[\Big(|\psi_1\rangle_{R_1C_1}+|\psi_3\rangle_{R_1C_1}\Big)\\\nonumber
&&
|01\rangle_{R_2}|11\rangle_{C_2}|1\rangle_{B_1}|00\rangle_{B_2}\\\nonumber
&&+|\psi_2\rangle_{R_1C_1}|01\rangle_{R_2}|11\rangle_{C_2}|0\rangle_{B_1}|10\rangle_{B_2}\\\nonumber
&&
+|\psi_4\rangle_{R_1C_1}|01\rangle_{R_2}|11\rangle_{C_2}|0\rangle_{B_1}|01\rangle_{B_2}\Big]\\\nonumber
&&+|g_2\rangle_{R_1C_1R_2C_2B_1B_2}
\end{eqnarray}
where $|\psi_i\rangle_{R_1C_1},\ i=1,2,3,4$ represents the first, second, third, fourth row of the target matrix, respectively:
\begin{eqnarray*}
|\psi_1\rangle_{R_1C_1}&=&\frac{1}{4}|00\rangle_{R_1}|00\rangle_{C_1}
+\frac{1}{16}|00\rangle_{R_1}|01\rangle_{C_1}\\\nonumber
&&+\frac{3}{16}|00\rangle_{R_1}|10\rangle_{C_1}
+\frac{3}{16}|00\rangle_{R_1}|11\rangle_{C_1},
\end{eqnarray*}
 \begin{eqnarray*}
|\psi_2\rangle_{R_1C_1}&=& \frac{3}{16}|01\rangle_{R_1}|00\rangle_{C_1}+\frac{3}{16}|01\rangle_{R_1}|01\rangle_{C_1}\\\nonumber
&&+\frac{2}{16}|01\rangle_{R_1}|10\rangle_{C_1}+\frac{2}{4}|01\rangle_{R_1}|11\rangle_{C_1},
 \end{eqnarray*}
   \begin{eqnarray*}
|\psi_3\rangle_{R_1C_1}&=&\frac{7}{16}|10\rangle_{R_1}|00\rangle_{C_1}
+0 |10\rangle_{R_1}|01\rangle_{C_1}\\\nonumber
&&+\frac{1}{4}|10\rangle_{R_1}|10\rangle_{C_1}
+0 |10\rangle_{R_1}|11\rangle_{C_1},
 \end{eqnarray*}
  \begin{eqnarray*}
|\psi_4\rangle_{R_1C_1}&=& 0|11\rangle_{R_1}|00\rangle_{C_1}+\frac{2}{4} |11\rangle_{R_1}|01\rangle_{C_1} \\\nonumber
&&+\frac{2}{16}|11\rangle_{R_1}|10\rangle_{C_1}+\frac{2}{16}|11\rangle_{R_1}|11\rangle_{C_1}
 \end{eqnarray*}

\textbf {Step 5: Label the useful information terms and useless
information terms}

Introduce a 1-qubit auxiliary $B_3$
in the state $|0\rangle_{B_3}$, and use $B_3$ to label the useful and useless information terms in $|\Phi_3\rangle$. To do this, construct a control operator:
$$W^{(4)}_{B_1B_2B_3}=V^{(1)}V^{(2)}V^{(3)},$$
where $V^{(i)}=P^{(i)}_{B_1B_2}\otimes \sigma^{(x)}_{B_3} + (I_{B_1B_2}-P_{B_1B_2}) \otimes I_{B_3}$,
$P^{(i)}_{B_1B_2}(i=1,2,3)$ is the projection operator
\begin{eqnarray*}
P^{(1)}_{B_1B_2}&=& |100\rangle_{B_1B_2} \;{_{B_1B_2}}\langle 100|\\\nonumber
P^{(2)}_{B_1B_2}&=& |001\rangle_{B_1B_2} \;{_{B_1B_2}}\langle 001|\\\nonumber
P^{(3)}_{B_1B_2}&=& |010\rangle_{B_1B_2} \;{_{B_1B_2}}\langle 010|.
\end{eqnarray*}
Apply $W^{(4)}_{B_1B_2B_3}$ to $|\Phi_3\rangle |0\rangle_{B_3}$, obtained:
\begin{eqnarray}
|\Phi_4\rangle&=& W^{(4)}_{B_1B_2B_3}|\Phi_3\rangle |0\rangle_{B_3} \\\nonumber
&=&\frac{1}{\sqrt{3}}\Big[\Big(|\psi_1\rangle_{R_1C_1}+|\psi_3\rangle_{R_1C_1}\Big)\\\nonumber
&&
|01\rangle_{R_2}|11\rangle_{C_2}|1\rangle_{B_1}|00\rangle_{B_2}\\\nonumber
&&+|\psi_2\rangle_{R_1C_1}|01\rangle_{R_2}|11\rangle_{C_2}|0\rangle_{B_1}|10\rangle_{B_2}\\\nonumber
&&
+|\psi_4\rangle_{R_1C_1}|01\rangle_{R_2}|11\rangle_{C_2}|0\rangle_{B_1}|01\rangle_{B_2}\Big]|1\rangle_{B_3}\\\nonumber
&&+|g_2\rangle_{R_1C_1R_2C_2B_1B_2}
\end{eqnarray}
Mark the useful information terms with $|1\rangle_{B_3}$ and the useless information terms with $|0\rangle_{B_3}$.

\textbf {Step 6: Use Hadamard gates for information superposition.}

Apply the Hadamard operator $W^{(5)}_{B_1B_2}=H^{\otimes 3}$ acts on $B_1$ and $B_2$,  obtained:
\begin{eqnarray}
|\Phi_5\rangle &=&W^{(5)}_{B_1B_2} |\Phi_4\rangle  \\\nonumber
&=&\frac{1}{(\sqrt{2})^{3}\cdot\sqrt{3}}\Big[|\psi_1\rangle_{R_1C_1}+|\psi_3\rangle_{R_1C_1}+|\psi_2\rangle_{R_1C_1}\\\nonumber
&&
+|\psi_4\rangle_{R_1C_1}\Big]|01\rangle_{R_2}|11\rangle_{C_2}|0\rangle_{B_1}|00\rangle_{B_2}|1\rangle_{B_3}\\\nonumber
&&+|g_3\rangle_{R_1C_1R_2C_2B_1B_2B_3}.
\end{eqnarray}

\textbf {Step 7: Measure the auxiliary $B_1,B_2,B_3$ in the state $|0001\rangle_{B_1B_2B_3}$.}

Finally, use $|0001\rangle_{B_1B_2B_3}$ to measure ancilla $B_1,B_2,B_3$, thus removing useless information items and obtaining useful information items as
\begin{eqnarray*}
|\Phi_6\rangle &=&
|\Psi_{out}\rangle\ |01\rangle_{R_2}|11\rangle_{C_2} ,
\end{eqnarray*}
where
\begin{eqnarray*}
|\Psi_{out}\rangle &=& |\psi_1\rangle_{R_1C_1}+|\psi_2\rangle_{R_1C_1}+|\psi_3\rangle_{R_1C_1}
+|\psi_4\rangle_{R_1C_1}
\end{eqnarray*}
The probability of the above measurement is $\frac{1}{24}$.


\begin{thebibliography}{99}
\bibitem{BP}  Benioff P. {\it The computer as a physical system: A microscopic quantum mechanical Hamiltonian model of computers as represented by Turing machines. } J. Stat. Phys., 1980, 22(5): 563­591.
\bibitem{MYI} Manin Y I. {\it Computable and noncomputable (in Russian)}. Sov. Radio, 1980, 13–15.
\bibitem{Deu} Deutsch D. {\it Quantum theory, the Church–Turing principle and the universal quantum computer. } \href{https://doi.org/10.1098/rspa.1985.0070}{ Proc. R. Soc. Lond. A, 1985, 400(1818): 97­117}.

\bibitem{Sho1} Shor P W.{\it Algorithms for quantum computation: discrete logarithms and factoring.} Proceedings 35th Annual Symposium on Foundations of Computer Science. \href{10.1109/SFCS.1994.365700}{IEEE, 1994: 124­134}.
\bibitem{Sho2} Shor P W. {\it Polynomial­time algorithms for prime factorization and discrete logarithms on a quantum computer. } \href{https://doi.org/10.1137/S0097539795293172}{SIAM Rev., 1999, 41(2): 303­332}.



\bibitem{Gro} Grover L K. {\it A fast quantum mechanical algorithm for database search.} \href{https://doi.org/10.1145/237814.237866}{Proceedings of the 28th Annual ACM Symposium on Theory of Computing. 1996: 212­219}.

\bibitem{HHL} Harrow A W, Hassidim A, Lloyd S. {\it Quantum algorithm for linear systems of equations.} \href{https://doi.org/10.1103/PhysRevLett.103.150502}{Phys. Rev. Lett., 2009, 103(15): 150502}.
\bibitem{HHL1} Clader B D, Jacobs B C, Sprouse C R. {\it Preconditioned quantum linear system algorithm. } \href{https://doi.org/10.1103/PhysRevLett.110.250504}{Phys. Rev. Lett., 2013, 110(25): 250504}.

\bibitem{QFT1} Coppersmith D, Winograd S.{\it Matrix multiplication via arithmetic progressions.} \href{https://api.semanticscholar.org/CorpusID:17450629}{J. Symb. Comput., 1990, 9(3): 251–280}.

\bibitem{QFT2} Weinstein Y S, Pravia M A, Fortunato E M, et al. {\it Implementation of the quantum Fourier transform. } \href{https://doi.org/10.1103/PhysRevLett.86.1889}{Phys. Rev. Lett., 2001, 86(9): 1889}.

\bibitem{QFT3} Nielsen M A, Chuang I. {\it Quantum Computation and Quantum Information.} Cambridge: Cambridge University Press, 2000.
\bibitem{Wang} Wang H, Wu L, Liu Y, et al. {\it Measurement­based quantum phase estimation algorithm for finding eigenvalues of non­unitary matrices. } \href{https://doi.org/10.1103/PhysRevA.82.062303}{Phys. Rev. A, 2010, 82(6):062303}.




\bibitem{MYM} Motta M, Ye E, McClean J R, et al. {\it Low rank representations for quantum simulation of electronic structure. \href{https://doi.org/10.1038/s41534-021-00416-z}{Npj Quantum Inf., 2021, 7: 83}.
\bibitem{KSG} Kockum A F, Soro A, García-Álvarez L, et al. {\it Lecture Notes on Quantum Computing. } \href{https://doi.org/10.48550/arXiv.2311.08445}{arXiv: 2311.08445 [quant-ph]}.

\bibitem{PA} Prakash A. {\it Quantum algorithms for linear algebra and machine learning.} \href{https://escholarship.org/uc/item/5v9535q4}{Ph.D. Thesis, UC Berkeley, 2014}.}

\bibitem{THF} Trotter H F. {\it On the product of semi-groups of operators.} P. Am. Math. Soc., 1959, 10(4): 545-551.

\bibitem{ZhaoZ} Zhao Z, Fitzsimons J K, Fitzsimons J F. {\it Quantum­assisted Gaussian process regression. } \href{https://doi.org/10.1103/PhysRevA.99.052331}{Phys. Rev. A, 2019, 99(5): 052331}.

\bibitem{SZ_2019} Stolze J, Zenchuk A I.{\it Computing scalar products via a two­terminal quantum transmission line. } \href{https://doi.org/10.1016/j.physleta.2019.125978}{Phys. Lett. A, 2019, 383(34): 125978}.


\bibitem{QZKW_arxive2022} Wentao Q, Alexander I Z, Asutosh K, Junde W,{\it Quantum algorithms for matrix operations and linear systems of equations. } \href{https://doi.org/10.1088/1572-9494/ad2366}{Commun. Theor. Phys., 2024, 76: 035103}.

\bibitem{ZQKW_arxiv2023} Alexander I Z, Wentao Q, Asutosh K, Junde W, {\it Matrix manipulations via unitary transformations and ancilla state measurements. } \href{https://doi.org/10.26421/QIC24.13-14-2}{Quantum Inf. Comput., 2024, 24(13) : 1099-1109}.


\bibitem{ZQKW_arxiv2024} Alexander I Z, Wentao Q, Asutosh K, Junde W, {\it Polynomialdepth quantum algorithm for computing matrix determinant.}
\href{https://doi.org/10.48550/arXiv.2401.16619}{arXiv: 2401.16619 [quant-ph]}.

\bibitem{ZQW_arxiv2025} Alexander I Z, Wentao Q, Junde W, {\it Quantum Hermitian conjugate and encoding unnormalized matrices.}
\href{https://doi.org/10.48550/arXiv.2504.00015}{arXiv:2504.00015 [quant-ph]}.

\bibitem{KShV}
 Kitaev A Y, Shen A H, Vyalyi M N, {\it Classical and Quantum Computation,} Graduate
Studies in Mathematics, V.47, American Mathematical Society, Providence, Rhode Island
(2002).

\end{thebibliography}
\end{document}